\documentclass[journal]{IEEEtran}
   \usepackage[pdftex]{graphicx}
  \DeclareGraphicsExtensions{.pdf,.jpeg,.png}
\usepackage[]{units}
\usepackage[T1]{fontenc} % optional
\usepackage{amsmath}
\usepackage{amssymb}
\usepackage{bm} % optional
\usepackage{eurosym}
\usepackage[super]{nth} % optional
\usepackage{multirow}
\usepackage{color}
\usepackage[caption=false,font=footnotesize]{subfig}
\usepackage{enumitem}
\usepackage{cite}

\usepackage{longtable} %It's here :)
\usepackage{eurosym}
\usepackage{comment}
\usepackage{float}

\usepackage[export]{adjustbox}
\usepackage[subpreambles=true]{standalone}

\usepackage{algorithm,caption}
\usepackage[table]{xcolor}

\begin{document}
%
% paper title
% Titles are generally capitalized except for words such as a, an, and, as,
% at, but, by, for, in, nor, of, on, or, the, to and up, which are usually
% not capitalized unless they are the first or last word of the title.
% Linebreaks \\ can be used within to get better formatting as desired.
% Do not put math or special symbols in the title.
\title{Extended Mathematical Derivations: Decentralized Model-free Loss Minimization in Distribution Grids with the Use of Inverters}
%
%
% author names and IEEE memberships
% note positions of commas and nonbreaking spaces ( ~ ) LaTeX will not break
% a structure at a ~ so this keeps an author's name from being broken across
% two lines.
% use \thanks{} to gain access to the first footnote area
% a separate \thanks must be used for each paragraph as LaTeX2e's \thanks
% was not built to handle multiple paragraphs
%

\author{Ilgiz~Murzakhanov,~\IEEEmembership{Student~Member,~IEEE,}
        Spyros~Chatzivasileiadis,~\IEEEmembership{Senior~Member,~IEEE}% <-this % stops a space
\thanks{This work is supported by the ID-EDGe project, funded by Innovation Fund Denmark, Grant Agreement No. 8127-00017B, and by the FLEXGRID project, funded by the European Commission Horizon 2020 program, Grant Agreement No. 863876.}% <-this % stops a space
\thanks{I. Murzakhanov and S. Chatzivasileiadis are with the Department of Wind and Energy Systems, Technical University of Denmark (DTU), Kgs. Lyngby, Denmark. E-mail: \{ilgmu, spchatz\}  @dtu.dk.}}

\maketitle

% As a general rule, do not put math, special symbols or citations
% in the abstract or keywords.
\begin{abstract}
This document contains extended mathematical derivations for the communication-free and model-free algorithms that can actively control converter-connected devices, and can operate either as stand-alone or in combination with centralized optimization algorithms. We address the problem of loss minimization in distribution grids, and we analytically prove that our proposed algorithms reduce the total grid losses without any prior information about the network, requiring no communication, and based only on local measurements. 
% .
\end{abstract}

% Note that keywords are not normally used for peerreview papers.

% For peer review papers, you can put extra information on the cover
% page as needed:
% \ifCLASSOPTIONpeerreview
% \begin{center} \bfseries EDICS Category: 3-BBND \end{center}
% \fi
%
% For peerreview papers, this IEEEtran command inserts a page break and
% creates the second title. It will be ignored for other modes.
\IEEEpeerreviewmaketitle

\section{Introduction}
% The very first letter is a 2 line initial drop letter followed
% by the rest of the first word in caps.
%
% form to use if the first word consists of a single letter:
% \IEEEPARstart{A}{demo} file is ....
%
% form to use if you need the single drop letter followed by
% normal text (unknown if ever used by the IEEE):
% \IEEEPARstart{A}{}demo file is ....
%
% Some journals put the first two words in caps:
% \IEEEPARstart{T}{his demo} file is ....
%
% Here we have the typical use of a "T" for an initial drop letter
% and "HIS" in caps to complete the first word.

\IEEEPARstart{M}{odern} distribution grids are characterized by the rapidly increasing penetration of distributed energy resources (DERs), especially photovoltaics (PVs) and battery storage systems. Reverse power flows and a greater ratio of fluctuating generation at the local level require the real-time efficient and secure operation of distribution grids. Considering the millions of DER units to be connected to the grid, however, it is almost impossible to manage their operation centrally in real-time. The computation and communication requirements for such a task go beyond the current capabilities of state-of-the art computation and communication infrastructure. Even if distributed algorithms are employed, it is improbable to have established a communication channel with all devices at all times. Parts of the grid will probably remain unobservable, or data will not be able to transmitted in real time. Therefore, communication-free (local) and model-free algorithms, which do not require any knowledge of the surrounding system are expected to play a significant role in the managing of such a system. Such algorithms, being agnostic to the topology of the system or the point where the device is connected, do not only offer plug'n'play capabilities, but, if designed appropriately, they can achieve system-wide objectives (e.g. optimal voltage profile, minimum losses, etc.) with local actions. In this document, we extend some mathematical derivations provided in the main paper. For compactness, during referencing to the main paper \cite{arxiv_main}, we use original numbers of the equations and denote equations of this extended document with an additional letter ``A''. 

The remainder of this document is organized as follows. First, we analytically compare losses during  the ``no-action'' strategy and  the local load measuring algorithm in Section~\ref{sec:NASvsLLMA}. In Section \ref{sec:LLMAvsLFMA}, we mathematically compare losses during  the local load and local flow measuring algorithms.

\section{Comparison of losses during  the ``no-action'' strategy and  the local load measuring algorithm} \label{sec:NASvsLLMA}
We claim that the local load measuring algorithm (LLMA) leads to lower or equal active power losses as the ``no-action'' strategy:
\begin{equation} \tag{1A} \label{NoAction_LocFlow_comp}
\Delta P^{{\mathcal{H}}} \leq  \Delta P^{\mathcal{N}}
\end{equation}

We prove \eqref{NoAction_LocFlow_comp} by comparing corresponding terms of $\Delta P^{{\mathcal{H}}}$ and $\Delta P^{\mathcal{N}}$. For compactness, we provide the comparison only for the first terms of $\Delta P^{{\mathcal{H}}}$ and $\Delta P^{\mathcal{N}}$, but a similar comparison can be done for other terms as well. 

According to (4b), the reactive power DistFlow equation for node $i$ looks as:
\begin{equation} \tag{2A} \label{DistFlow_Q_i}
Q_{i} + Q_{i'} = Q_{0} - x_{0}\frac{P^2_{0} + (Q_{0})^2}{V_0^2} - Q^L_{i} + Q^G_{i}
\end{equation}

As reactive generation setpoints are 0 for  the ``no-action'' strategy, \eqref{DistFlow_Q_i} transforms to:
\begin{equation} \tag{3A} \label{NoAct_Q_i}
Q^{\mathcal{N}}_{i} + Q^{\mathcal{N}}_{i'} = Q^{\mathcal{N}}_{0} - x_{0}\frac{P^2_{0} + (Q^{\mathcal{N}}_{0})^2}{V_0^2} - Q^L_{i}
\end{equation}

Note that the line reactance $x_{0}$, PV output $P^2_{0}$, reactive load $Q^L_{i}$, voltage magnitude of a slack bus $V_0$ do not change across considered loss minimization algorithms. As a result, the algorithm superscript for them is omitted. 

Similarly, equation \eqref{DistFlow_Q_i} for LLMA is as follows:
\begin{equation} \tag{4A} \label{LoadMeas_Q_i}
Q^{\mathcal{H}}_{i} + Q^{\mathcal{H}}_{i'} = Q^{\mathcal{H}}_{0} - x_{0}\frac{P^2_{0} + (Q^{\mathcal{H}}_{0})^2}{V_0^2} - Q^L_{i} + Q^{G,{\mathcal{H}}}_{i}
\end{equation}
By solving a quadratic equation \eqref{LoadMeas_Q_i} with respect to $Q^{\mathcal{H}}_{0}$, we obtain two solutions:
\begin{equation} \tag{5A} \label{Q0_sol_i}
Q^{\mathcal{H}}_{0} = (1 \pm \sqrt{D^{\mathcal{H}}}) \frac{V_0^2}{2x_0}
\end{equation}
where $D^{\mathcal{H}}$ is defined as:
\begin{equation} \tag{6A} \label{D_i}
D^{\mathcal{H}} = 1 - 4 \frac{x_0}{V_0^2} (\frac{x_0 P_0^2}{V_0^2} + Q_i^{\mathcal{H}} + Q_{i'}^{\mathcal{H}} + Q_i^L) + 4\frac{x_0}{V_0^2} Q_i^{G,{\mathcal{H}}}
\end{equation}
Note, that \eqref{Q0_sol_i} has only one physically feasible solution:
\begin{equation} \tag{7A} \label{Q0_feas_i}
Q^{\mathcal{H}}_{0} = (1 - \sqrt{D^{\mathcal{H}}}) \frac{V_0^2}{2x_0}
\end{equation}
% The latter is also obvious from the fact that $D \approx 1$, so using $1 + \sqrt{D}$ would result in $Q^{\mathcal{H}}_{0}=\frac{V_0^2}{2x_0}$. But this is only possible if $V_i = 0$. % That depends on P_0, Q_L, Q^G_{max} values. The greater P_0 and greater a difference between Q_L and Q^G_{max}, the smaller \sqrt{D} from 1.
Obviously, when $Q_i^{G,{\mathcal{H}}} = 0$, then equations \eqref{DistFlow_Q_i} and \eqref{LoadMeas_Q_i} become identical. As a result, the values of $Q^{\mathcal{H}}_{0}$ are also the same:
\begin{equation} \tag{8A} \label{Q0_Eq}
Q^{\mathcal{N}}_{0} = Q^{\mathcal{H}}_{0} | Q_i^{G,{\mathcal{H}}} = 0
\end{equation}
For the case of LLMA, increasing $Q_i^{G,{\mathcal{H}}}>0$ leads to increase of $D^{\mathcal{H}}$, defined by \eqref{D_i}. And in accordance with \eqref{Q0_feas_i}, it results in lower $Q^{\mathcal{H}}_{0}$ value. Note that change of $Q_i^{G,{\mathcal{H}}}>0$ does not affect flows $Q_i^{\mathcal{H}}$ and $Q_{i'}^{\mathcal{H}}$, as the corresponding branches are located downstream from bus $i$. 

Summing up, we have shown that increasing generation $Q_i^{G}$ in node $i$ leads to lower power flow $Q_0$:%, while flows $Q_i$ and $Q_{i'}$ remain the same. 
\begin{equation} \tag{9A} \label{Q0_comp1}
Q^{\mathcal{H}}_{0} < Q^{\mathcal{N}}_{0}  
\end{equation}
Then it is straightforward that for the first term of (6) the following relation holds:
\begin{equation} \tag{10A} \label{Q0_comp2}
r_0\frac{P^2_0 + (Q_0^{\mathcal{H}})^2}{V^2_0} < r_0\frac{P^2_0 + (Q_0^\mathcal{N})^2}{V^2_0}
\end{equation}
By providing similar proofs for other terms of (6), we conclude that:
\begin{equation} \tag{11A} \label{NoAct_LocLoad_Fin}
\begin{cases} \Delta P^{\mathcal{H}} = \Delta P^\mathcal{N}, & \mbox{if all } Q^{G,{\mathcal{H}}} = 0
\\ \Delta P^{\mathcal{H}} < \Delta P^\mathcal{N}, & \mbox{otherwise} \end{cases}
\end{equation}

Note that provided proof can be adapted for any radial system by varying a number of terms in (5).

\section{Comparison of losses during  the local load and local flow measuring algorithms} \label{sec:LLMAvsLFMA}
We claim that the local flow measuring algorithm (LFMA) provides lower or equal active power losses as the local load measuring algorithm (LLMA):
\begin{equation} \tag{12A} \label{LocLoad_LocFlow_comp}
\Delta P^{{\mathcal{F}}} \leq \Delta P^{{\mathcal{H}}}
\end{equation}

Similarly to Section \ref{sec:NASvsLLMA}, we prove \eqref{LocLoad_LocFlow_comp} by comparing only the first terms of $\Delta P^{{\mathcal{F}}}$ and $\Delta P^{{\mathcal{H}}}$. By conducting the similar derivations for LFMA as in \eqref{LoadMeas_Q_i}-\eqref{Q0_feas_i}, we obtain:
\begin{equation} \tag{13A} \label{Q0_feasF_i}
Q^{\mathcal{F}}_{0} = (1 - \sqrt{D^{\mathcal{F}}}) \frac{V_0^2}{2x_0}
\end{equation}
where $D^{\mathcal{F}}$ is given as follows:
\begin{equation} \tag{14A} \label{DF_i}
D^{\mathcal{F}} = 1 - 4 \frac{x_0}{V_0^2} (\frac{x_0 P_0^2}{V_0^2} + Q_i^{\mathcal{F}} + Q_{i'}^{\mathcal{H}} + Q_i^L) + 4\frac{x_0}{V_0^2} Q_i^{G,{\mathcal{F}}}
\end{equation}

Note that flow $Q_{i'}$ does not change between the steps of Algorithm 2 and remains equal to $Q^{\mathcal{H}}_{i'}$, as node $i'+1$ is a leaf node. The only different terms between \eqref{D_i} and \eqref{DF_i} are power flows $Q_i$ and generation setpoints $Q^G_i$. As step 4 of Algorithm 2 consists of several cases, we prove \eqref{LocLoad_LocFlow_comp} by considering all the cases.

\subsection{LFMA-step 4: upstream flow does not change direction}
In this section we prove that \eqref{LocLoad_LocFlow_comp} holds if upstream flow does not change direction between steps 2 and 3. As a result, only steps 1-3 of Algorithm 2 are performed. Then generation setpoint of bus $i$ is:
\begin{equation} \tag{15A} \label{eq:busi}
\begin{cases} Q_i^{G,{\mathcal{F}}} = Q_i^{G,{\mathcal{I}}} = Q_i^{G,{\mathcal{H}}}, & \mbox{if (2)-(3) are binding for} 
\\  & \mbox{inverter $i$ in step 2} 
\\ Q_i^{G,{\mathcal{F}}} = Q_i^{G,{\mathcal{I}}} > Q_i^{G,{\mathcal{H}}}, & \mbox{otherwise} \end{cases}
\end{equation}
Equations similar to \eqref{eq:busi} can be written for bus $i+1$, which imply the following relations for upstream flow $i$:
\begin{equation} \tag{16A} \label{eq:busi1}
\begin{cases} Q_i^{\mathcal{F}} = Q_i^{\mathcal{I}} = Q_i^{\mathcal{H}}, & \mbox{if (2)-(3) are binding for} 
\\  & \mbox{inverter $i+1$ in step 2} 
\\ Q_i^{\mathcal{F}} = Q_i^{\mathcal{I}} < Q_i^{\mathcal{H}}, & \mbox{otherwise} \end{cases}
\end{equation}
Considering \eqref{eq:busi} and \eqref{eq:busi1}, we define the relation between $D^{\mathcal{F}}$, which is defined in \eqref{DF_i}, and $D^{\mathcal{H}}$, which is defined in \eqref{D_i}:
\begin{equation} \tag{17A} \label{eq:compFH}
\begin{cases} D^{\mathcal{F}} = D^{\mathcal{H}}, & \mbox{if (2)-(3) are binding for} 
\\  & \mbox{inverters $i$ and $i+1$ in step 2} 
\\ D^{\mathcal{F}} > D^{\mathcal{H}}, & \mbox{otherwise} \end{cases}
\end{equation}
Next, with the use of determined \eqref{eq:compFH} we compare $Q^{\mathcal{F}}_{0}$, which is defined in \eqref{Q0_feasF_i}, and $Q^{\mathcal{H}}_{0}$, which is defined in \eqref{Q0_feas_i}:
\begin{equation} \tag{18A} \label{eq:compQ0}
\begin{cases} Q^{\mathcal{F}}_{0} = Q^{\mathcal{H}}_{0}, & \mbox{if (2)-(3) are binding for} 
\\  & \mbox{inverters $i$ and $i+1$ in step 2} 
\\ Q^{\mathcal{F}}_{0} < Q^{\mathcal{H}}_{0}, & \mbox{otherwise} \end{cases}
\end{equation}
Finally, with \eqref{eq:compQ0} it is easy to compare the first term of (6) for LFMA and LLMA, in the same manner as it is done in \eqref{Q0_comp2}. By providing similar proofs for other terms of (6), we conclude:
\begin{equation} \tag{19A} \label{eq:compLP}
\begin{cases} \Delta P^{{\mathcal{F}}} = \Delta P^{{\mathcal{H}}}, & \mbox{if (2)-(3) are binding for} 
\\  & \mbox{inverters $i$ and $i+1$ in step 2} 
\\ \Delta P^{{\mathcal{F}}} < \Delta P^{{\mathcal{H}}}, & \mbox{otherwise} \end{cases}
\end{equation}
Next, we consider two cases when the upstream flow changes direction between steps 2 and 3.

\subsection{LFMA-step 4: downstream flow does not change direction}
In this section, we prove \eqref{LocLoad_LocFlow_comp} for a case when downstream flow does not change direction between steps 2 and 3.

Note that downstream flow $Q_{i+1}$ of a bus $i+1$ cannot change its own direction between steps 2 and 3, as bus $i+2$ is a leaf node. Thus, the only bus in the considered system, which can experience a direction change of its own downstream flow, is node $i$. Generation setpoint of bus $i$ after step 2 is defined by (7), while after step 4 is determined by (9) and (8). Note that inverter limits (2)-(3) are not binding in (7) and (8) for the considered case, as then upstream flow $Q_0$ would not change its own direction, which contradicts to made assumption. Thus, generation setpoints of bus $i$ after steps 2 and 4 are:
\begin{subequations}
\begin{alignat}{2}
& Q_i^{G,{\mathcal{H}}} = Q_i^L \tag{20Aa} \label{Qi_step1}\\
&  Q_i^{G,{\mathcal{F}}} = Q_i^L + Q_i^{\mathcal{H}} - |Q^{\mathcal{I}}_{0}| \tag{20Ab} \label{Qi_step3} 
\end{alignat}
\end{subequations}
After subtracting \eqref{Qi_step1} from \eqref{Qi_step3} we get:
\begin{equation} \tag{21A} \label{eq:subtr}
Q_i^{G,{\mathcal{F}}} - Q_i^{G,{\mathcal{H}}} = Q_i^{\mathcal{H}} - |Q^{\mathcal{I}}_{0}| 
\end{equation}
As step 3 leads to higher local generation, then power flows in lines connected to non-leaf nodes decrease, so:
\begin{equation} \tag{22A} \label{eq:Comp_flow}
Q^{\mathcal{H}}_{i} - Q^{\mathcal{I}}_{0} > 0 
\end{equation}
As the right part of \eqref{eq:subtr} is positive due to \eqref{eq:Comp_flow}, so is the left part, then:
\begin{equation} \tag{23A} \label{eq:Qg_comp}
Q_i^{G,{\mathcal{F}}} > Q_i^{G,{\mathcal{H}}}
\end{equation}
By deriving relations for $Q_i$, $D$, $Q_0$ variables in LLMA and LFMA, similarly to \eqref{eq:busi}-\eqref{eq:compLP}, we conclude that: 
\begin{equation} \tag{24A} \label{eq:upComp}
\Delta P^{\mathcal{F}} < \Delta P^{\mathcal{H}}
\end{equation}

\subsection{LFMA-step 4: downstream flow changes direction}
In this section we prove \eqref{LocLoad_LocFlow_comp} for a case when downstream flow changes direction between steps 2 and 3, so generation $Q_i^{G,{\mathcal{F}}}$ is set according to (7). As in this case $Q_i^{G,{\mathcal{F}}} = Q_i^{G,{\mathcal{H}}}$, by following derivations similar to \eqref{eq:busi}-\eqref{eq:compLP}, we obtain:
\begin{equation} \tag{25A} \label{eq:upComp2}
\Delta P^{\mathcal{F}} = \Delta P^{\mathcal{H}}
\end{equation}

\subsection{Summary of LLMA and LFMA comparison}
Taking into account \eqref{eq:compLP}, \eqref{eq:upComp}, \eqref{eq:upComp2}, we conclude that:
\begin{equation} \tag{26A} \label{eq:upComp3}
\Delta P^{\mathcal{F}} \leq \Delta P^{\mathcal{H}}
\end{equation}
where equality is achieved only if (2)-(3) are binding or downstream flow changes direction between steps 2 and 3.

Note that provided proof can be adapted for any radial system by varying a number of terms in (5).

\bibliographystyle{IEEEtran}
\bibliography{library}
 %%%%%%%%%%%%%%%%%%%%%%%%%%%%%%%%%%%%%%%%%%%%%%%%%%%%%%%%%%%%%%%%
%
 %%%%%%%%%%%%%%%%%%%%%%%%%%%%%%%%%%%%%%%%%%%%%%%%%%%%%%%%%%%%%%%%
 \end{document}